\documentclass[10pt]{article}

\usepackage{amsmath}
\usepackage{amssymb}

\usepackage{graphicx}

\usepackage{cite}

\usepackage{color}

\usepackage{float}

\usepackage[labelfont=bf,labelsep=period,justification=raggedright]{caption}

\makeatletter
\renewcommand{\@biblabel}[1]{\quad#1.}
\makeatother

\date{}

\begin{document}

\begin{flushleft}
{\Large
\textbf{Agent-based model with asymmetric trading and herding for complex financial systems}
}
\\
Jun-Jie Chen,
Bo Zheng$^{\ast}$,
Lei Tan
\\
Department of Physics, Zhejiang University, Hangzhou, Zhejiang, China
\\
$\ast$ E-mail: zheng@zimp.zju.edu.cn
\end{flushleft}

\section*{Abstract}

\emph{\textbf{Background:}} For complex financial systems, the negative and positive return-volatility correlations, i.e., the so-called leverage and anti-leverage effects, are particularly important for the understanding of the price dynamics. However, the microscopic origination of the leverage and anti-leverage effects is still not understood, and how to produce these effects in agent-based modeling remains open. On the other hand, in constructing microscopic models, it is a promising conception to determine model parameters from empirical data rather than from statistical fitting of the results.
\\
\\
\emph{\textbf{Methods:}} To study the microscopic origination of the return-volatility
correlation in financial systems, we take into account the individual and collective behaviors of investors in real markets, and construct an agent-based model.
The agents are linked with each other and trade in groups, and particularly, two novel microscopic mechanisms, i.e.,
investors' asymmetric trading and herding in bull and bear markets, are introduced. Further, we propose effective methods to determine the key parameters in our model from historical market data.
\\
\\
\emph{\textbf{Results:}} With the model parameters determined for six representative stock-market indices in the world respectively, we obtain the corresponding leverage or anti-leverage effect from the simulation, and the effect is in agreement with the empirical one on amplitude and duration. At the same time, our model produces other features of the real markets, such as the fat-tail distribution of returns and the long-term correlation of volatilities.
\\
\\
\emph{\textbf{Conclusions:}} We reveal that for the leverage and anti-leverage effects, both the investors' asymmetric trading and herding are essential generation mechanisms. Among the six markets, however, the investors' trading is approximately symmetric for the five markets which exhibit the leverage effect, thus contributing very little. These two microscopic mechanisms and the methods for the determination of the key parameters can be applied to other complex systems with similar asymmetries.


\section*{Introduction}

In recent years, the understanding of complex systems has been undergoing rapid development. Financial markets are important examples of complex systems with
many-body interactions. The possibility of accessing large amounts
of historical financial data has spurred the interest of
scientists in various fields, including physics.
Plenty of results have been obtained with physical concepts, methods and models \cite{man95,gop99,liu99,bou01,gab03,qiu06,she09,qiu10,zha11,pre11,zho11,jia12,jia13}.

There are several stylized facts in
financial markets. Besides the fat tail in the probability distribution of price returns,
it is well-known that the volatilities are long-range
correlated in time, which is the so-called volatility clustering \cite{yam05}. However, our knowledge on the dynamics of the price itself is still limited. Since the auto-correlation of returns is extremely weak \cite{gop99,liu99}, nonzero higher-order time correlations become important, especially the lowest-order one among them. In financial markets, this lowest-order nonzero correlation turns out to be the return-volatility correlation, on which we lay emphasis in this paper. In 1976, a negative return-volatility correlation is first discovered by Black\cite{bla76}. This is the so-called leverage effect, which implies that past negative returns increase future volatilities. The leverage effect is actually observed in various financial systems, such as stock markets, futures markets, bank interest rates and foreign exchange rates \cite{bla76,eng01,bol06,qiu06,qiu07,she09a,par11,pre12}. We have studied about thirty stock-market indices, and all of them exhibit the leverage effect. To the best of our knowledge, the leverage effect exists in almost all stock markets in the world.
In Chinese stock markets, however, a positive return-volatility
correlation is detected, which is called the anti-leverage effect \cite{qiu06,she09a}. This effect is also observed in other economic systems, such as bank interest rates of early years and spot markets of non-ferrous metals.

The leverage and anti-leverage effects are crucial for the understanding of the price dynamics \cite{bla76,qiu06,she09a,par11}, and important for risk management and optimal portfolio choice \cite{bou01a,bur10}. However, the origination of the return-volatility correlation is still disputed, even at the macroscopic level \cite{hau91,bek00,gir04,ahl07,rom08,she09a,par11,li11}. According to Black, the leverage
effect arises because a price drop increases the risk of a company to
go bankrupt and leads the stock to fluctuate more. So far, various
macroscopic models have been proposed to understand the return-volatility correlation \cite{bai96,bou01,tan06,mas06,rui08,she09a}. The retarded volatility model
is an enlightening one, which can produce both the
leverage and anti-leverage effects
\cite{bou01}. However, it is a model with only one degree of freedom, and both the initial time series of returns and the function of the
feedback return-volatility interaction, are actually input. Hence, the model is phenomenological in essence, and the
generation mechanism of the leverage and anti-leverage effects is macroscopic. In very recent years, many researches have been devoted to the return-volatility correlation, but how to produce the return-volatility correlation with a microscopic
model remains open.

Agent-based modeling is a powerful simulation technique, which is widely applied in various fields \cite{gia01,cha01,bon02,eba04,ren06a,far09,sch09,fen12}. More recently, an agent-based
model is proposed for reproducing the cumulative distribution of empirical returns
and trades in stock markets\cite{fen12}. It is a outstanding model with key parameters
determined from empirical findings rather than from being set artificially. In this paper, we construct
an agent-based model with asymmetric trading and herding to explore the
microscopic origination of the leverage and anti-leverage effects. In the past decades, although the asymmetric trading and herding behaviors may have been touched macroscopically, they have not been taken into account in the microscopic modeling yet. Especially, we propose effective methods to determine the key parameters in our model from historical market data.

\section*{Methods}

To study the microscopic origination of the return-volatility correlation in stock markets, we take into account the individual and collective behaviors of investors, and construct a microscopic model with multi-agent interactions. Further, we determine the key parameters in our model from historical market data rather than from statistical fitting of the results.

Our model is basically built on agents' daily trading, i.e., buying, selling and holding stocks. Empirical studies indicate that investors make decisions according to the previous stock performance
of different time windows \cite{men10}, which suggests
that their horizons of investment vary. This investment horizon is introduced
to our model for a better description of agents' market behavior.
Most crucially, two important behaviors of investors are taken into account for understanding the return-volatility correlation.

\textbf{1. Two important behaviors of investors}

(a) Investors' asymmetric trading in bull and bear markets. There are various definitions of bull and bear markets \cite{pag02,jan10}. The usual definition is that in stock markets, bull and bear markets correspond to the periods of generally
increasing and decreasing stock prices respectively \cite{pag02}. In this paper we adopt this definition, and simply define a market to be bullish on one day if the price return is positive,
and bearish if the price return is negative. The asymmetric trading in bull and bear markets is an individual behavior, which is induced by investors' different trading desire when the price drops and rises. To be more specific, an investor's willingness to trade is
affected by the previous price returns, leading the trading probability to be distinct in bull and bear markets.

(b) Investors' asymmetric herding in bull and bear markets. Herding, as one of the collective behaviors, is that investors cluster in groups when making decisions, and these groups can be large in financial markets \cite{egu00,con00,hwa04,zhe04,ken11,ken12,ken13}. Actually, the herding behavior in bull markets is not the same as that in bear ones \cite{hwa04,kim05,wal06}. For instance, previous study has shown that in the recent
US market, the herding behavior in bear markets appears
much more significant than that in bull ones \cite{hwa04}. Generally, investors may
cluster more intensively in either bull or bear markets, leading the herding to be asymmetric.

\textbf{2. Microscopic model with multi-agent interactions}

The stock price on day $t$ is denoted as $Y(t)$, and the logarithmic price return is $R(t)=\ln [Y(t)/Y(t-1)]$. In stock markets,
the information for investors is highly incomplete, therefore
an agent's decision of \emph{buy}, \emph{sell} or \emph{hold} is assumed to be random. Since intraday trading is not persistent in empirical trading
data \cite{eis07}, we consider that only one trading decision is made by each agent in a single day. In our model, there are $N$ agents, and each operates one share every day.
On day $t$, each agent $i$ makes a
trading decision $S_{i}(t)$,
\begin{equation}
S_{i}(t)=\begin{cases}
\:1 & \textnormal{buy}\\
-1 & \textnormal{sell}\\
\:0 & \textnormal{hold}
\end{cases}\label{eq:si},
\end{equation}
and the probabilities of
buy, sell and hold decisions are denoted as $P_{buy}(t)$, $P_{sell}(t)$ and $P_{hold}(t)$, respectively.
The price return $R(t)$ in our model
is defined by the difference of the demand and supply of the stock, i.e.,
the difference between the number of buy agents
and sell ones,
\begin{equation}
R(t)=\sum_{i=1}^{N}S_{i}(t).\label{eq:return}
\end{equation}
The volatility is defined as the absolute return $|R(t)|$.

The investment horizon is introduced since agents' decision makings are based on the previous stock performance
of different time horizons. It has been found
that the relative portion $\gamma_{i}$ of agents with $i$ days investment
horizon follows a power-law decay, $\gamma_{i}\varpropto i^{- \eta}$
with $\eta=1.12$ \cite{fen12}. The maximum investment horizon is denoted as $M$, thus $i_{max}=M$. With the condition of $\sum_{i=1}^{M}\gamma_{i}=1$,
we normalize $\gamma_i$ to be $\gamma_{i}=i^{-\eta}/\sum_{i=1}^{M}i^{-\eta}$. Agents' trading decisions are made according to the previous price returns. For an agent having investment
horizon of $i$ days, $\sum_{j=0}^{i-1}R(t-j)$ represents a simplified
investment basis for decision making on day $t+1$. We introduce a weighted average return $R'(t)$
to describe the integrated investment basis of all agents. Taking into account that $\gamma_{i}$ is the weight of $\sum_{j=0}^{i-1}R(t-j)$, $R'(t)$ is defined as
\begin{equation}
R'(t)=k\cdot\sum_{i=1}^{M}\left[\gamma_{i}\sum_{j=0}^{i-1}R(t-j)\right],\label{eq:fR}
\end{equation}
where $k$ is a proportional coefficient. We set $k=1/(\sum_{i=1}^{M}\sum_{j=i}^{M}\gamma_{j})$, such that $|R'(t)|_{max}=N=|R(t)|_{max}$ to ensure that the fluctuation scale of $R'(t)$ remains consistent with the one of $R(t)$ (see Appendix S1). If $M=1$, $R'(t)$ is just identical to $R(t)$.
Actually, $M$ varies from market to market, and from time period to time period for a market. According to Ref. \cite{men10}, the investment horizons of investors range from a few days to several months. We estimate the maximum investment horizon $M$ to be $150$ in our model. For $M$ between $50$ and $500$, the simulated results remain qualitatively robust.

\textbf{(i) Asymmetric trading.} In Ref. \cite{fen12}, investors' probabilities of buy and sell are assumed to be equal, i.e., $P_{buy}=P_{sell}=p$, and $p$ is a constant. In our model, we adopt the value of $p$ estimated in Ref. \cite{fen12}, $p=0.0154$. We assume $P_{buy}(t)=P_{sell}(t)$ as well, but now $P_{buy}(t)$ and $P_{sell}(t)$ evolve with time since the agents' trading is asymmetric in bull and bear markets. As the trading probability $P_{trade}(t)=P_{buy}(t)+P_{sell}(t)$, we set its average over time $\langle P_{trade}(t)\rangle =2p$. From the investors' behavior (a) described in Subsec. 1 in Sect. Methods, we define the market performance of the
previous $M$ days to be bullish if $R'(t)>0$,
and bearish if $R'(t)<0$.
The investors' asymmetric trading in bull and bear markets
gives rise to the distinction between $P_{trade}(t+1)|_{R'(t)>0}$ and $P_{trade}(t+1)|_{R'(t)<0}$.
Thus, $P_{trade}(t+1)$ should take the form
\begin{equation}
\begin{cases}
P_{trade}(t+1)=2p\cdot\alpha & \; R'(t)>0\\
P_{trade}(t+1)=2p & \; R'(t)=0\\
P_{trade}(t+1)=2p\cdot\beta & \; R'(t)<0
\end{cases}\label{eq:fp}.
\end{equation}
Here $\alpha$ and $\beta$ are constants, and $\langle P_{trade}(t)\rangle =2p$ requires $\alpha+\beta=2$, i.e., $\alpha$ and $\beta$ are not independent.

\textbf{(ii) Asymmetric herding.} The herding behavior implies that investors can be divided into groups. Here a herding degree $D(t)$ is introduced to quantify the clustering degree of the herding behavior,
\begin{equation}
D(t)=n_A(t)/N,\label{eq:dt}
\end{equation}
where $n_A(t)$ is the average number of agents in each group on day $t$. Herding should be related to previous volatilities \cite{con00,bla12}, and we set $n_A(t+1)=|R'(t)|$. Hence
the herding degree on day $t+1$ is
\begin{equation}
D(t+1)=|R'(t)|/N.\label{eq:od}
\end{equation}
This herding degree is symmetric for $R'(t)>0$ and $R'(t)<0$.
According to the investors' behavior (b) described in Subsec. 1 in Sect. Methods, however, investors' herding behaviors in bull and bear markets are asymmetric, i.e., herding is stronger in either bull markets or bear ones. More specifically, $D(t+1)$ is not symmetric for $R'(t)>0$ and $R'(t)<0$, and should be redefined to
be
\begin{equation}
D(t+1)=|R'(t)-\Delta R|/N.\label{eq:fd}
\end{equation}
Here $\Delta R$ is the degree of asymmetry, and as $\Delta R$ grows, herding becomes more asymmetric. According to Eq.~(\ref{eq:dt}), $D(t+1)=n_A(t+1)/N$. Therefore $N\cdot D(t+1)$ is the average number of agents in a same group. Thus we randomly divide $N$ agents into
$1/D(t+1)$ groups on day $t+1$. Everyday, the agents in a group make a same trading decision
(buy, sell or hold) with the same probability ($P_{buy}$, $P_{sell}$
or $P_{hold}$).

\textbf{3. Determination of $\alpha$ and $\Delta R$}

This is the key step in the construction of our model. We emphasize that $\alpha$ and $\Delta R$ are determined from the historical market data rather than from statistical fitting of the simulated results.
Six representative stock-market indices are studied with our model, including the S\&P 500, Shanghai, Nikkei 225, FTSE 100, Hangseng and DAX indices. We collect the daily data of closing price and trading volume, both of which are from 1950 to 2012 with 15775 data points for the S\&P 500
Index, from 1991 to 2006 with 3928 data points for the Shanghai Index, from 2003 to 2012 with 2367 data points for the Nikkei 225 Index, from 2004 to 2012 with 1801 data points for the FTSE 100 Index, from 2001 to 2012 with 2787 data points for the Hangseng Index
and from 2008 to 2012 with 1016 data points for the DAX Index. These data are obtained from "Yahoo$!$ Finance" (http://finance.yahoo.com).
For comparison of different time series of returns, the normalized
return $r(t)$ is introduced,
\begin{equation}
r(t)=[R(t)-\langle R(t)\rangle]/\sigma,\label{eq:norm}
\end{equation}
where $\langle \cdots\rangle $ represents the average
over time $t$, and $\sigma=\sqrt{\langle R^{2}(t)\rangle -\langle R(t)\rangle ^{2}}$ is the standard deviation of
$R(t)$.

The stock market is assumed to be bullish if $r(t)>0$,
and bearish if $r(t)<0$. To determine $\alpha$, we first define an average trading volume $V_{+}$
for the bull markets, and $V_{-}$ for the bear ones,
\begin{equation}
\left\{ \begin{array}{c}
V_{+}=[\sum_{r(t)>0}V(t)]/n_{r(t)>0}\\
V_{-}=[\sum_{r(t)<0}V(t)]/n_{r(t)<0}
\end{array}\right..
\end{equation}
Here $n_{r(t)>0}$ and $n_{r(t)<0}$ represent the number of positive and negative returns respectively, and $V(t)$ is the trading volume on day $t$.
As displayed in Table~\ref{tab:value}, the ratio $V_{+}/V_{-}$ is $1.03$ for the S\&P 500 Index and $1.21$ for the Shanghai Index.
In our model, since the average trading volumes for bull markets ($R'(t)>0$) and bear markets ($R'(t)<0$) are $N\cdot P_{trade}(t+1)|_{R'(t)>0}$ and $N\cdot P_{trade}(t+1)|_{R'(t)<0}$,
the ratio of these two average trading volumes is
\begin{equation}
\frac{P_{trade}(t+1)|_{R'(t)>0}}{P_{trade}(t+1)|_{R'(t)<0}}=\alpha/\beta=V_{+}/V_{-}.
\end{equation}
Together with the condition $\alpha+\beta=2$, we determine $\alpha=1.01$ from $V_{+}/V_{-}$ for the S\&P 500 Index and $\alpha=1.09$ for the Shanghai Index.
Table~\ref{tab:value} also shows the values of $V_{+}/V_{-}$ and $\alpha$ for the Nikkei 225, FTSE 100, Hangseng and DAX indices. Several data series of different time periods are sampled from the historical market data, and the error is given for $\alpha$ in this table. Student's \emph{t}-test is performed to analyze the statistical significance for $\alpha$ deviating from $1.0$, and a \emph{p}-value less than 0.05 is considered statistically significant. The analysis shows that only the value $\alpha=1.09$ of the Shanghai Index is significantly deviating from $1.0$, with the $\textnormal{\emph{p}-value}=8.4\times10^{-4}$. In our simulation, for simplicity, we approximate $\alpha$ to be $1.0$ for the S\&P 500, Nikkei 225, FTSE 100, Hangseng and DAX indices, and $1.1$ for the Shanghai Index.

Now we turn to $\Delta R$.
In real markets, herding is related to volatilities \cite{con00,bla12}. Thus we introduce the average $|r(t)|$ with the weight $V(t)$ to describe the herding degree in a specific period. Thus the herding degrees of bull markets ($r(t)>0$) and bear markets ($r(t)<0$) are defined as
\begin{equation}
\left\{ \begin{array}{c}
d_{bull}(r(t))=\sum_{t,r(t)>0}[V(t)\cdot r(t)]/\sum_{t,r(t)>0}V(t)\\
d_{bear}(r(t))=\sum_{t,r(t)<0}[V(t)\cdot|r(t)|]/\sum_{t,r(t)<0}V(t)
\end{array}\right.\label{eq:reald}.
\end{equation}
From empirical findings, the herding degrees of bull and bear stock markets are not equal, i.e., $d_{bull}\neq d_{bear}$.
In order to equalize $d_{bull}$ and $d_{bear}$, we introduce a shifting to $r(t)$, denoted by $\Delta r$, such that $d_{bull}(r'(t))=d_{bear}(r'(t))$ with $r'(t)=r(t)+\Delta r$. From this  definition of $\Delta r$, we derive (see Appendix S2)
\begin{equation}
\Delta r=\frac{1}{2}[d_{bear}(r(t))-d_{bull}(r(t))].
\end{equation}
Thus we obtain $\Delta r=0.067$ for the S\&P 500 Index and $\Delta r=-0.043$ for the Shanghai Index. In our model, we similarly compute
the shifting to the time series $R(t)$, which equalize the herding degree $D(t+1)=|R'(t)-\Delta R|/N$
in bull markets ($R'(t)>0$) and bear markets ($R'(t)<0$). Actually, one may prove that the shifting to $R(t)$ is equivalent to the shifting to $R'(t)$ (see Appendix S3). If $R'(t)$ is replaced by $R''(t)=R'(t)+\Delta R$,
$D(t+1)$ turns into $D(t+1)=|R''(t)-\Delta R|/N=|R'(t)|/N$,
which is symmetric for bull and bear markets. Therefore, $\Delta R$ is the shifting to $R'(t)$, and it is just the shifting to $R(t)$.

\begin{figure}[H]
\begin{center}
\includegraphics[width=4in]{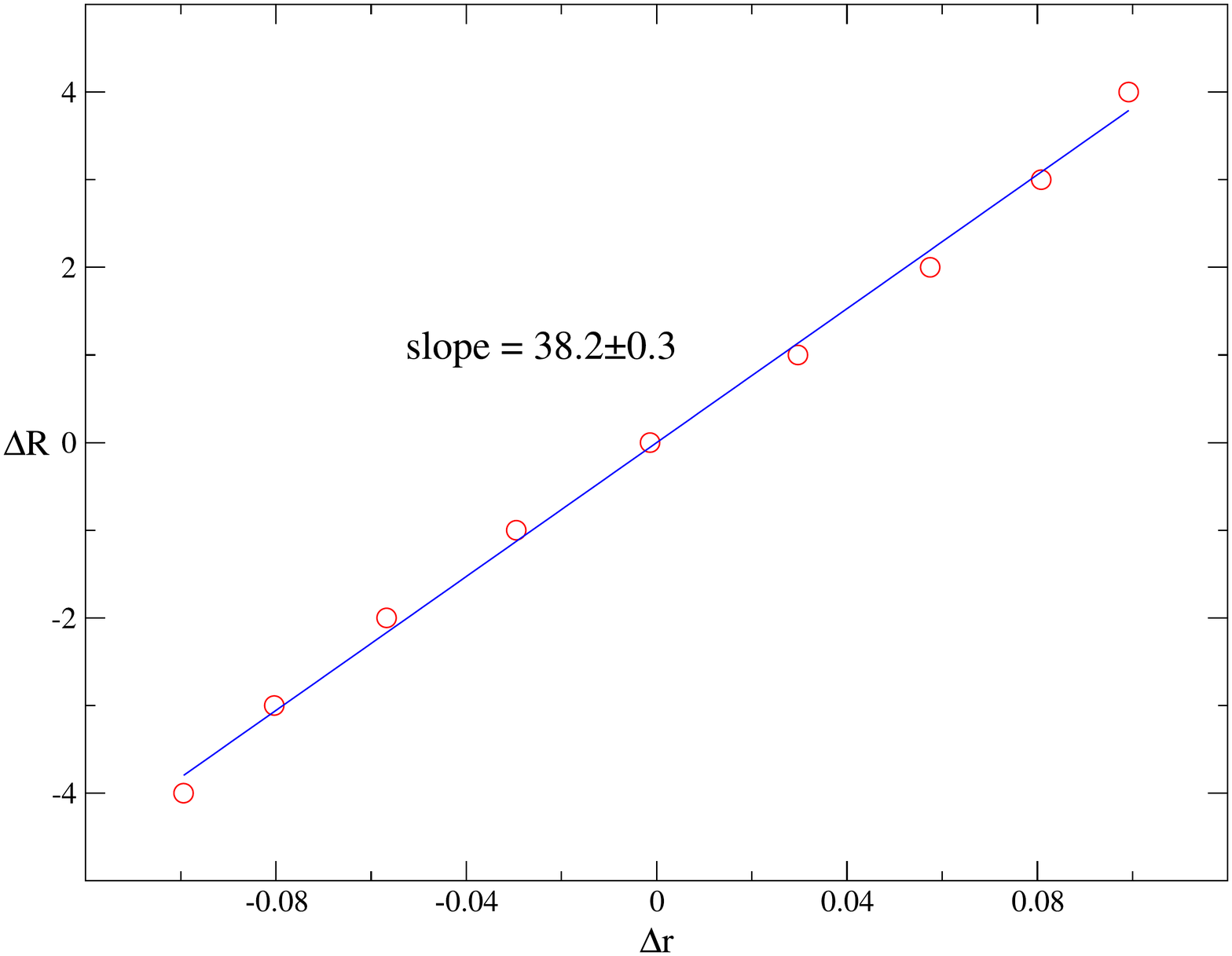}
\end{center}
\caption{\textbf{The relation of $\Delta R$ and $\Delta r$.
} With $\Delta R$ set to be $-4$, $-3$, $-2$, $-1$, $0$, $1$, $2$, $3$ and $4$ respectively, time series $R(t)$ is simulated $100$ times for $\alpha=1.0$. The corresponding $\Delta r$ is computed and averaged for each $\Delta R$. This plot shows a linear relation of $\Delta R$ and $\Delta r$, i.e., $\Delta
R=38.2\Delta r$, and this result remains robust for $\alpha$ between $0.9$ and
$1.1$.
}
\label{fig:Delatr}
\end{figure}

The time series of returns in different real markets and in our model fluctuate at different levels. For comparison, we normalize the returns with Eq.~(\ref{eq:norm}). Similarly, $\Delta R$, the shifting to returns, should also be normalized to $\Delta r$. However, in simulating the stock markets with our model, the parameter we need is $\Delta R$. Therefore, we should first derive the relation of $\Delta R$ and $\Delta r$. With the normalization of the time series $R(t)$, $\Delta R$ should be normalized to $\Delta r$,
\begin{equation}
[\Delta R-\langle R(t)\rangle]/\sigma=\Delta r,
\end{equation}
where $\langle \cdots\rangle $ represents the average
over time $t$, and
$\sigma$ is the standard deviation of $R(t)$. To determine the relation of $\Delta R$ and $\Delta r$, $\Delta R$ is set
to be $-4$, $-3$, $-2$, $-1$, $0$, $1$, $2$, $3$, $4$, respectively, and $\alpha$ is set to be $1.0$ to produce time series $R(t)$. With $R(t)$
simulated $100$ times for each $\Delta R$, we compute $\Delta r$
and average the results. As displayed in Fig.~\ref{fig:Delatr}, the relation of $\Delta R$ and $\Delta r$ is linear, and $\Delta
R=38.2\Delta r$.
For $\alpha$ between $0.9$ and $1.1$, the results remain robust. Thus,
we determine $\Delta R=3$ for the simulation of the S\&P 500 Index and $\Delta R=-2$ for the simulation of the Shanghai Index. Table~\ref{tab:value} shows the values of $\Delta r$ and $\Delta R$, as well as the error of $\Delta r$,  for the Nikkei 225, FTSE 100, Hangseng and DAX indices. Due to the fluctuation of the empirical data, the error of $\Delta r$ is about $10$ percent. Since the sign of $\Delta r$ determines that the simulation yields the leverage or anti-leverage effect, we perform Student's \emph{t}-test to analyze the statistical significance of $\Delta r$, and the corresponding \emph{p}-value is listed in Table~\ref{tab:value}. A \emph{p}-value less than 0.05 is considered statistically significant.

To further validate the methods for the determination of the key parameters and the simulations for the leverage and anti-leverage effects, eight more indices are studied (see Appendix S4). The simulation of each index correctly produces the leverage or anti-leverage effect.

\textbf{4. Simulation}

The number of agents
in our simulations is $10000$, i.e., $N=10000$. With $\alpha$ and $\Delta R$ determined for each index, our model produces the time
series of returns $R(t)$ in the following procedure. Initially, the returns of the first $150$ time steps are set to be 0. On day $t+1$, we calculate $R'(t)$ according to Eq.~(\ref{eq:fR}), then $P_{trade}(t+1)$ and $D(t+1)$ according to Eq.~(\ref{eq:fp}) and Eq.~(\ref{eq:fd}), respectively. Next, we randomly divide all agents into
$1/D(t+1)$ groups. The agents in a group make a same trading decision (buy, sell or hold) with the same probability ($P_{buy}$, $P_{sell}$
or $P_{hold}$). After all agents have made their decisions, we calculate the return $R(t+1)$ with Eq.~(\ref{eq:si}) and Eq.~(\ref{eq:return}). Repeating this procedure, we obtain the return time series $R(t)$. $20000$ data points of $R(t)$ are produced in each simulation, but the first $10000$ data points are abandoned for equilibration.

\section*{Results}
To describe how past returns affect future volatilities, the return-volatility correlation function $L(t)$ is defined,
\begin{equation}
L(t)=[\langle r(t')\cdot|r(t'+t)|^{2}\rangle -L_{0}] /Z,
\end{equation}
with $Z=\langle |r(t')|^{2}\rangle ^{2}$
and $L_{0}=\langle r(t')\rangle \langle |r(t')|^{2}\rangle $
\cite{bou01}. Here $\langle \cdots\rangle $ represents the average
over time $t'$.

\begin{figure}[H]
\begin{center}
\includegraphics[width=4in]{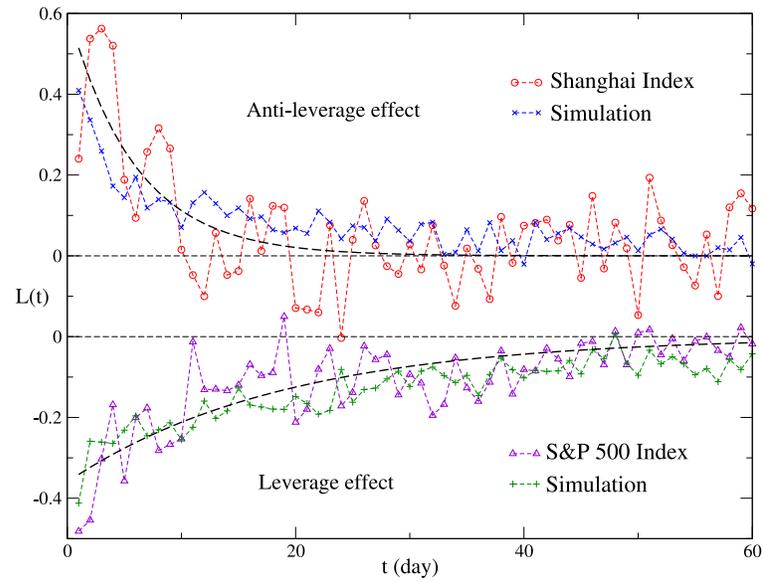}
\end{center}
\caption{\textbf{The return-volatility correlation functions for the S\&P 500 and Shanghai indices, and for the corresponding simulations.} The S\&P 500 and Shanghai indices are simulated
with $(\alpha,\Delta R)=(1.0,3)$ and $(\alpha,\Delta R)=(1.1,-2)$, respectively. Dashed lines show an exponential fit $L(t)=c\cdot exp(-t/\tau)$
with $(c,\tau)=(-0.36,19)$ and $(0.61,8)$ for the S\&P 500 Index and the
Shanghai Index.
}
\label{fig:L}
\end{figure}

As displayed in Fig.~\ref{fig:L}, $L(t)$ calculated with the empirical data of the S\&P 500 Index shows negative values up to at least 15 days, and this is the well-known leverage effect \cite{bou01,bla76,qiu06}. On the other hand, $L(t)$ for the Shanghai Index remains positive for about 10 days. That is the so-called anti-leverage effect \cite{qiu06,she09a}. Fitting $L(t)$ to an exponential form $L(t)=c\cdot exp(-t/\tau)$, we obtains $\tau=19$ and $8$ days for the leverage and anti-leverage effects, respectively. Compared with the short correlating time of the returns, the order of minutes \cite{gop99,liu99}, both the leverage and anti-leverage effects are prominent. As the lowest-order nonzero correlations of returns, the leverage and anti-leverage effects are theoretically crucial for the understanding of the price dynamics \cite{bla76,qiu06,she09a,par11}. In practical application, these effects are important for risk management and optimal portfolio choice \cite{bou01a,bur10}. After the time series $R(t)$ produced in our model is normalized
to $r(t)$, we compute the return-volatility correlation function, and the result is in agreement with that calculated from empirical data on amplitude and duration for both the S\&P 500 and Shanghai indices, as shown in Fig.~\ref{fig:L}. This is the first time that the leverage and anti-leverage effects are produced with a microscopic model.

\begin{figure}[H]
\begin{center}
\includegraphics[width=4in]{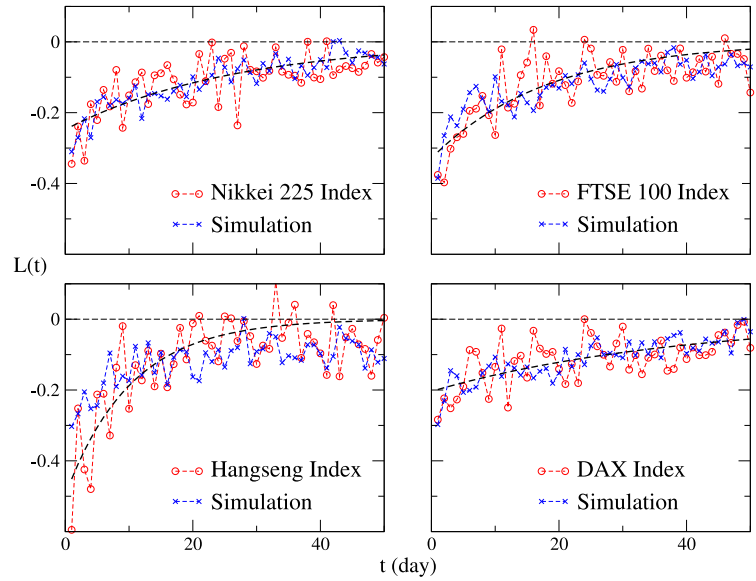}
\end{center}
\caption{\textbf{The return-volatility correlation functions for the four indices and the corresponding simulations.
} The Nikkei 225, FTSE 100, Hangseng and DAX indices are simulated
with $(\alpha,\Delta R)=(1.0,2)$, $(1.0,2)$, $(1.0,2)$ and $(1.0,1)$, respectively. Dashed lines show an exponential fit $L(t)=c\cdot exp(-t/\tau)$
with $(c,\tau)=(-0.25,26)$ for the Nikkei 225 Index, $(-0.33,18)$ for the FTSE 100 Index, $(-0.50,10)$ for the Hangseng Index and $(-0.20,39)$ for the DAX Index.
}
\label{fig:FandNL}
\end{figure}

For the Nikkei, FTSE 100, Hangseng and DAX indices, the volume data of early years are not available to us. However, $L(t)$ is computed from only price data. In order to reduce the fluctuation of $L(t)$, we collect the price data of a longer period, which are from 1984 to 2012 with 7092 data points for the Nikkei 225 Index, from
1984 to 2012 with 7227 data points for the FTSE 100 Index, from 1988 to
2012 with 6181 data points for the Hangseng Index and from 1990 to 2012
with 5514 data points for the DAX Index.
As displayed in Fig.~\ref{fig:FandNL}, $L(t)$ for the simulations is in agreement with that for the corresponding indices.
Table~\ref{tab:test} shows the values of $c$ and $\xi$ of the exponential fit $L(t)=c\cdot exp(\xi t)$ for the six indices and the corresponding simulations. Since $c$ is obviously non-zero, the \emph{p}-value of Student's \emph{t}-test is only listed for~$\xi$.

\begin{figure}[H]
\begin{center}
\includegraphics[width=4in]{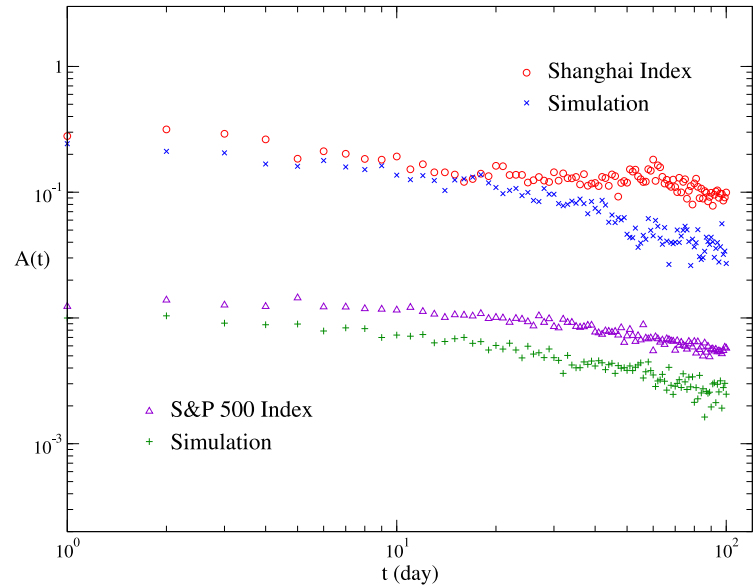}
\end{center}
\caption{\textbf{The auto-correlation functions of volatilities for
the S\&P 500 and Shanghai indices, and for the corresponding simulations.} For clarity, the curves for the S\&P 500 Index
have been shifted down by a factor of 10.
}
\label{fig:A}
\end{figure}

\begin{figure}[H]
\begin{center}
\includegraphics[width=4in]{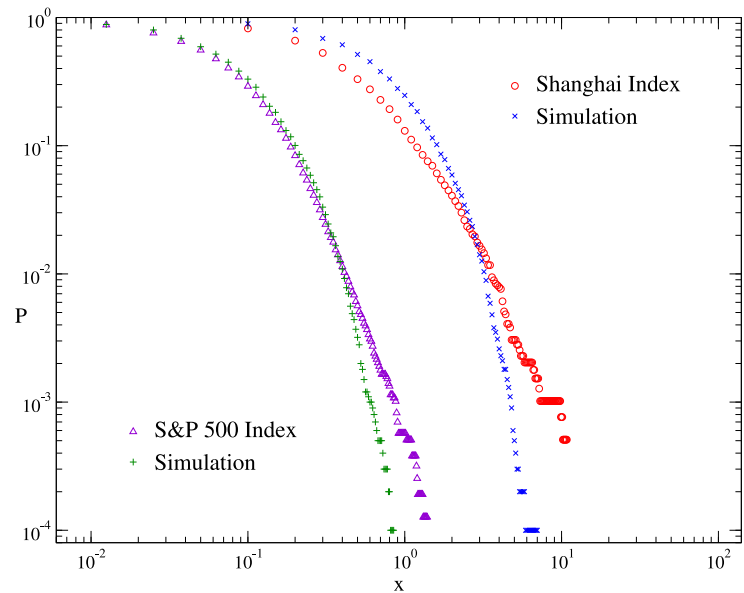}
\end{center}
\caption{\textbf{The cumulative distributions of absolute returns
for the S\&P 500 and Shanghai indices, and for the corresponding simulations.}
For clarity, the curves for the S\&P 500 Index
have been shifted left by a factor of 8.5.
}
\label{fig:P}
\end{figure}

Our model also produces other features of the real markets, such as the long-term correlation of volatilities and the fat-tail distribution of the returns. Here we take the S\&P 500 and Shanghai indices as examples. The auto-correlation function of volatilities is defined as
\begin{equation}
A(t)=[\langle |r(t')||r(t'+t)|\rangle -\langle |r(t')|\rangle ^{2}]/A_{0},
\end{equation}
where $A_{0}=\langle |r(t')|^{2}\rangle -\langle |r(t')|\rangle ^{2}$ \cite{she09a}, and $\langle \cdots\rangle $ represents the average
over time $t'$. As shown in Fig.~\ref{fig:A}, $A(t)$ for the simulations is consistent with that for the empirical data.
The cumulative distributions $P(|r(t)|>x)$ of absolute returns are shown in Fig.~\ref{fig:P}, where the fat tail in the distribution of empirical returns can be observed in that of the simulated returns as well.

By the definitions, both $\alpha$ and $\Delta r$ are not dependent on the number of agents (denoted by $N$) in the model. However, the slope of the linear relation between $\Delta R$ and $\Delta r$ increases with $N$. Therefore, the magnitude of $\Delta R$ becomes larger as $N$ grows. For the simulation results, the amplitude of $L(t)$ increases with $N$, but gradually converges for larger $N$ (see Appendix S5). For $A(t)$ and $P(|r(t)|>x)$, the cases are similar.

\section*{Discussion}
In our model, the crucial generation mechanisms of the return-volatility correlation are the agents' asymmetric trading
and herding behaviors in bull and bear markets. Now we discuss how these two mechanisms contribute to the leverage and anti-leverage effects, and which one is more significant. According to Eq.~(\ref{eq:fp}) and $\alpha+ \beta=2$, $P_{trade}$ is symmetric about $R'(t)=0$ if $\alpha=1.0$, and asymmetric if $\alpha \neq 1.0$.  On the other hand, $D(t+1)$ in Eq.~(\ref{eq:fd}) is asymmetric about $R'(t)=0$ if $\Delta R \neq 0$. In our model, the S\&P 500 and Shanghai indices are simulated with$(\alpha,\Delta R)=(1.0,3)$ and $(\alpha,\Delta R)=(1.1,-2)$, respectively. Therefore, $P_{trade}$ is symmetric in the simulation of the S\&P 500 Index, but asymmetric in the simulation of the Shanghai Index. $D(t+1)$ is asymmetric in the simulation of both the S\&P 500 and Shanghai indices. With other parts of the model remain unchanged, we consider the following controls: (a) $P_{trade}$ is replaced by a symmetric one in the simulation of the Shanghai Index; (b) $D(t+1)$ is replaced by a symmetric one in the simulation of both the S\&P 500 and Shanghai indices; (c) both $P_{trade}$ and $D(t+1)$ are replaced by the symmetric ones in the simulation of the Shanghai Index.

\begin{figure}[H]
\begin{center}
\includegraphics[width=4in]{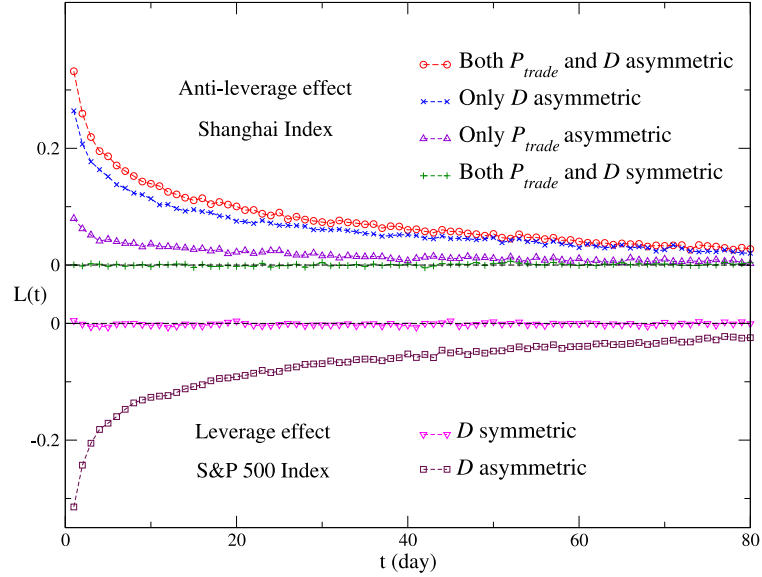}
\end{center}
\caption{\textbf{The return-volatility correlation functions for the simulated results of the S\&P 500 and Shanghai indices, and for those of the controls.} The S\&P 500 and Shanghai indices exhibit the leverage and anti-leverage effects, respectively. For the leverage effect, we consider two cases: $D$ is asymmetric; $D$ is symmetric. The latter is the control. For the anti-leverage effect, we consider the following cases: both $P_{trade}$ and $D$ are asymmetric; only $D$ is asymmetric; only $P_{trade}$ is asymmetric; both $P_{trade}$ and $D$ are symmetric. The last three cases are controls. For each case, the simulation is performed for $100$ times, and the average $L(t)$ is displayed.
}
\label{fig:SPSHL}
\end{figure}

The simulations are performed 100 times for average. We conclude that for the leverage and anti-leverage effects, both the investors' asymmetric trading and herding are essential generation mechanisms. As displayed in Fig.~\ref{fig:SPSHL}, the anti-leverage effect is weakened significantly
and the leverage effect disappears after we replace the asymmetric $D(t+1)$ with the symmetric one. On the other hand, the anti-leverage effect recedes after the asymmetric $P_{trade}$ is replaced by the symmetric one. It is worth mentioning that for the five stock markets exhibiting the leverage effect, the S\&P 500, Nikkei 225, FTSE 100, Hangseng and DAX,
$P_{trade}$ is approximately symmetric, thus contributing very little to the leverage effect.
The investors' asymmetric
trading in the Shanghai market may result from the fact that the Shanghai market is an emerging market. Investors are somewhat speculative, and rush
for trading as the stock price increases \cite{qiu06}.

\textbf{Conclusion}

Based on investors' individual and collective behaviors, we construct an agent-based model to investigate how the return-volatility correlation
arises in stock markets. In our model, agents are linked with each other and trade in groups. In
particular, two novel mechanisms, investors'
asymmetric trading and herding behaviors in bull and bear markets, are introduced. There are four parameters in our model, i.e., $p$, $M$, $\alpha$ and $\Delta R$. We adopt $p$ estimated in Ref. \cite{fen12}, and estimate the only tunable parameter $M$ to be $150$. $\alpha$ and $\Delta R$, the key parameters, are induced by the asymmetries in trading and herding, respectively. Specifically, we determine $\alpha$ from the ratio of the average trading volume when stock price is rising and that when price is dropping, and $\Delta R$ from investors' different herding degrees in bull and bear markets.

We collect the daily price and volume data of six representative stock-market indices in the world, including the S\&P 500, Shanghai, Nikkei 225, FTSE 100, Hangseng and DAX indices. With $\alpha$ and $\Delta R$ determined for these indices respectively, we obtain the corresponding leverage or anti-leverage
effect from the simulation, and the effect is in agreement with
the empirical one on amplitude and duration. Other features, such as the long-range auto-correlation of volatilities and the fat-tail distribution
of returns, are produced at the same time. Further, it is quantitatively demonstrated in our model that both the investors' asymmetric trading and herding are essential generation mechanisms for the leverage and anti-leverage
effects at the microscopic level. However, the investors' trading is approximately symmetric for the five stock markets exhibiting the leverage effect, thus contributing very little to the effect. These two microscopic mechanisms and the methods for the determination of $\alpha$ and $\Delta R$ can also be applied to other complex economic systems with similar asymmetries in individual and collective behaviors, e.g., to futures markets, bank interest rates, foreign exchange rates and spot markets of non-ferrous metals.

\section*{Supporting Information}
\noindent\textbf{Appendix S1} Derivation of $k$

\noindent(PDF)
~\\

\noindent\textbf{Appendix S2} Derivation of $\Delta r$

\noindent(PDF)
~\\

\noindent\textbf{Appendix S3} Equivalence of the shifting to $R(t)$ and that to $R'(t)$

\noindent(PDF)
~\\

\noindent\textbf{Appendix S4} The values of $\alpha$, $\Delta r$ and $\Delta R$
for eight more indices

\noindent(PDF)
~\\

\noindent\textbf{Appendix S5} How $N$ affects the model parameters and simulation results

\noindent(PDF)

\section*{Acknowledgments}

\bibliography{AsymmetricAgentbasedModel.bbl}

\section*{Figure Legends}

\section*{Tables}

\begin{table}[H]
\caption{\textbf{The values of $V_{+}/V_{-}$, $d_{bull}$, $d_{bear}$, $\alpha$, $\Delta r$ and $\Delta R$
for the six indices.} $V_{+}/V_{-}$, $d_{bull}$ and $d_{bear}$ are determined
from the historical data for each index. We calculate $\alpha$ from  $\alpha+ \beta=2$ and $\alpha/\beta=V_{+}/V_{-}$, and $\Delta r$ from $\Delta r = \frac{1}{2}(d_{bear}-d_{bull})$. Student's \emph{t}-test is performed to analyze the statistical significance of $\Delta r$. A \emph{p}-value less than 0.05 is considered statistically significant. We compute $\Delta R$ from the linear relation between $\Delta r$ and $\Delta R$ for all these indices. As $\Delta R$ for the Shanghai Index is negative, it is rounded down to the nearest integer, while $\Delta R$ for other indices are positive, and each of them is rounded up to the nearest integer.
}
\begin{tabular}{|c|ccc|cccc|}
\hline
Index & $V_{+}/V_{-}$ & $d_{bull}$ & $d_{bear}$ & $\alpha$ & $\Delta r$ & \textit{p}-value & $\Delta R$\tabularnewline
\hline
S\&P 500 (1950-2012) & $1.03$ & $0.993$ & $1.127$ & $1.01\pm0.01$ & $0.067\pm0.007$ & $6.7\times10^{-4}$ & $3$\tabularnewline
Shanghai (1991-2006) & $1.21$ & $0.533$ & $0.447$ & $1.09\pm0.01$ & $-0.043\pm0.005$ & $1.0\times10^{-3}$ & $-2$\tabularnewline
Nikkei 225 (2003-2012) & $1.01$ & $0.729$ & $0.807$ & $1.01\pm0.01$ & $0.039\pm0.005$ & $1.5\times10^{-3}$ & $2$\tabularnewline
FTSE 100 (2004-2012) & $0.98$ & $0.673$ & $0.729$ & $0.99\pm0.01$ & $0.028\pm0.003$ & $7.3\times10^{-4}$ & $2$\tabularnewline
Hangseng (2001-2012) & $1.04$ & $0.966$ & $1.029$ & $1.02\pm0.02$ & $0.032\pm0.003$ & $4.4\times10^{-4}$ & $2$\tabularnewline
DAX (2008-2012) & $0.96$ & $0.797$ & $0.822$ & $0.98\pm0.02$ & $0.013\pm0.002$ & $2.9\times10^{-3}$ & $1$\tabularnewline
\hline
\end{tabular}
\label{tab:value}
\end{table}

\begin{table}[H]
\caption{\textbf{The values of $c$ and $\xi$ of the exponential fit $L(t)=c\cdot exp(\xi t)$ for the six indices and the corresponding simulations.} Student's \emph{t}-test is performed to analyze the statistical significance of $\xi$. A \emph{p}-value less than 0.05 is considered statistically significant.
}
\begin{tabular}{|c|c|cc|}
\hline
 & $c$ & $\xi$ & \textit{p}-value\tabularnewline
\hline
S\&P 500 & $-0.36\pm0.02$ & $-0.053\pm0.005$ & $4.5\times10^{-4}$\tabularnewline
simulation & $-0.30\pm0.01$ & $-0.032\pm0.001$ & $5.7\times10^{-6}$\tabularnewline
\hline
Shanghai & $0.61\pm0.12$ & $-0.133\pm0.014$ & $6.9\times10^{-4}$\tabularnewline
simulation & $0.30\pm0.02$ & $-0.066\pm0.004$ & $7.9\times10^{-5}$\tabularnewline
\hline
Nikkei 225 & $-0.25\pm0.01$ & $-0.038\pm0.004$ & $6.9\times10^{-4}$\tabularnewline
simulation & $-0.27\pm0.01$ & $-0.042\pm0.001$ & $1.9\times10^{-6}$\tabularnewline
\hline
FTSE 100 & $-0.33\pm0.03$ & $-0.055\pm0.007$ & $1.4\times10^{-3}$\tabularnewline
simulation & $-0.26\pm0.01$ & $-0.036\pm0.001$ & $3.6\times10^{-6}$\tabularnewline
\hline
Hangseng & $-0.50\pm0.06$ & $-0.098\pm0.012$ & $1.2\times10^{-3}$\tabularnewline
simulation & $-0.22\pm0.01$ & $-0.027\pm0.001$ & $1.1\times10^{-5}$\tabularnewline
\hline
DAX & $-0.20\pm0.01$ & $-0.026\pm0.002$ & $2.0\times10^{-4}$\tabularnewline
Simulation & $-0.22\pm0.01$ & $-0.031\pm0.001$ & $6.5\times10^{-6}$\tabularnewline
\hline
\end{tabular}
\label{tab:test}
\end{table}

\end{document}